\RequirePackage{fix-cm}
\documentclass[smallextended]{svjour3}       % onecolumn (second format)
\smartqed  % flush right qed marks, e.g. at end of proof
\usepackage{graphicx}
\usepackage{latexsym,amsfonts,bm,amsmath,amsbsy,bbm,epsfig,mathrsfs}
\usepackage{natbib}
\journalname{Journal of Mathematical Biology}

\newcommand{\dtau}{\mbox{d}\tau}

\newcommand{\du}{\mbox{d}u}
\newcommand{\dv}{\mbox{d}v}
\newcommand{\dz}{\mbox{d}z}

\newcommand{\dxi}{\mbox{d}\xi}
\newcommand{\sign}{\mathop{\rm sign}}
\begin{document}

\title{Coarse-graining molecular dynamics: stochastic models with non-Gaussian 
force distributions}

\titlerunning{Coarse-graining molecular dynamics}        % if too long for running head

\author{Radek Erban}

\institute{Radek Erban \at
Mathematical Institute, University of Oxford,
Radcliffe Observatory Quarter, Woodstock Road, Oxford, OX2 6GG, 
United Kingdom \\
\email{erban@maths.ox.ac.uk} 
}

\date{{\it Dedicated to Professor Hans Othmer on the occasion of his $75$th birthday.}}

\maketitle

\begin{abstract}
Incorporating atomistic and molecular information into models of cellular behaviour
is challenging because of a vast separation of spatial and temporal scales between 
processes happening at the atomic and cellular levels. Multiscale or 
multi-resolution methodologies address this difficulty by using molecular 
dynamics (MD) and coarse-grained models in different parts of the cell. 
Their applicability depends on the accuracy and properties of the 
coarse-grained model which approximates the detailed MD description. A family 
of stochastic coarse-grained (SCG) models, written as relatively low-dimensional
systems of nonlinear stochastic differential equations, is presented.
The nonlinear SCG model incorporates the non-Gaussian force distribution 
which is observed in MD simulations and which cannot be described by linear 
models. It is shown that the nonlinearities can be chosen in such a way
that they do not complicate parametrization of the SCG description by detailed MD 
simulations. The solution of the SCG model is found in terms of gamma functions. 
\keywords{multiscale modelling \and coarse-graining \and molecular dynamics \and Brownian dynamics}
\end{abstract}

\section{Introduction}
\label{secintro}

With increased experimental information on atomic or near-atomic
structure of bio\-molecules and intracellular components, there has 
been a growing need to incorporate such microscopic data (coming 
from X-ray crystallography, NMR spectroscopy or cryo-electron 
microscopy) into dynamical models of intracellular processes.
A common approach is to use molecular dynamics (MD) simulations 
based on classical molecular mechanics. Such MD models 
are written as relatively large systems of ordinary or stochastic 
differential equations for the positions and velocities of individual 
atoms, which can also be subject to algebraic 
constraints~\citep{Leimkuhler:2015:MD,Lewars:2016:CC}.
Although all-atom MD simulations of systems consisting of a million of 
atoms have been reported in the 
literature~\citep{Tarasova:2017:AMD,Farafanov:2019:MBC}, such
simulations are restricted to relatively small computational domains,
which are up to tens of nanometres long. It is beyond the 
reach of state-of-the-art computers to simulate intracellular processes 
which include transport of molecules over micrometers, because this would 
require simulations of trillions of atoms~\citep{Erban:2019:SMR}.

An example is modelling of calcium (Ca$^{2+}$) dynamics. On one hand, 
at the macroscopic level, Ca$^{2+}$ waves can propagate between cells 
over hundreds of micrometres and
\citet{Kang:2009:SCC} developed a model of Ca$^{2+}$ waves in a network of 
astrocytes. It builds on previous modelling work by \citet{Kang:2007:VCS}
describing intracellular Ca$^{2+}$ dynamics as a system of differential equations 
for concentrations of chemical species involved, including 
inositol 1,4,5-trisphosphate (IP$_3$), a chemical 
signal that binds to the IP$_3$ receptor to release Ca$^{2+}$ ions from the 
endoplasmic reticulum. On the other hand, at the atomic level, \citet{Hamada:2017:IGM} recently
solved IP$_3$-bound and unbound structures of large cytosolic domains of the IP$_3$
receptor by X-ray crystallography and clarified the IP$_3$-dependent gating mechanism 
through a unique leaflet structure.
 
Although it is not possible to incorporate such a detailed information
into Ca$^{2+}$ modelling by using  all-atom MD in the entire 
intracellular space, there is still potential to design
multiscale (multi-resolution) models which compute Ca$^{2+}$ 
dynamics with the resolution of individual Ca$^{2+}$ ions.
\citet{Dobramysl:2016:PMM} implement such a methodology at the Brownian
dynamics (BD) level to study Ca$^{2+}$ puff statistics stemming from 
IP$_3$ receptor channels. Denoting the position of an individual Ca$^{2+}$ ion by
${\mathbf X} \equiv (X_1,X_2,X_3)$, its diffusive BD trajectory is
given by
\begin{equation}
\mbox{d}X_i = \sqrt{2 D} \; \mbox{d}W_i, 
\qquad
\mbox{for}
\;\;
i=1,2,3,
\label{BDSDE}
\end{equation}
where $D$ is the diffusion constant and $W_i,$ $i=1,2,3,$ are three 
independent Wiener processes. Since 
individual positions of Ca$^{2+}$ ions are only needed in the 
vicinity of channel sites, \citet{Dobramysl:2016:PMM} model diffusion of ions far away
of the channel by a coarser model, utilizing the two-regime method 
developed by \citet{Flegg:2012:TRM}. This method enables efficient
simulations with the BD level of resolution by coarse-graining
the BD model in those parts of the simulation domain, where the
coarse-grained model can be safely used without introducing significant
numerical errors~\citep{Flegg:2014:ATM,Flegg:2015:CMC,Robinson:2015:MRS}. 

Although BD models or their multi-resolution extensions simulate 
individual molecules of chemical species involved, the binding of Ca$^{2+}$ ions 
to channel sites or other interactions between molecules are only 
described using relatively coarse probabilistic approaches. For example,
the BD model of \citet{Dobramysl:2016:PMM} describes interactions
in terms of reaction radii and binding probabilities as implemented 
by~\citet{Erban:2009:SMR} and \citet{Lipkova:2011:ABD}. 
Atomic-level information is not included in BD models. 
In order to use this information, multi-resolution methodologies have 
to consider MD simulations in parts of the simulation
domain. In the case of ions, such a multi-resolution scheme has been
developed by~\citet{Erban:2016:CAM}, where an all-atom MD model of ions
in water is coupled with a stochastic coarse-grained~(SCG) description 
of ions in the rest of the computational domain. 

The accuracy and efficiency of such multi-resolution methodologies 
depend on the quality of the SCG description of the 
underlying MD model. In this paper, we present and analyze a class of SCG
models which can be used to fit non-Gaussian distributions estimated from
all-atom MD simulations. While the velocity distribution of the
coarse-grained particle can be well approximated by a Gaussian (normal)
distribution in our MD simulations, this is not the case of the force 
distribution. Non-Gaussian force distributions have also been reported 
by~\cite{Shin:2010:BMM} and \cite{Carof:2014:TAC} in their MD simulations
of particles in Lennard-Jones fluids. Thus our SCG model is formulated
in a way which incorporates a Gaussian distribution for
the velocity and a non-Gaussian distribution for the force (acceleration).

Given an integer $N \ge 1$,
a coarse-grained particle (for example, an ion) will be
described by $(2N+2)$ three-dimensional variables: its position
${\mathbf X}$, velocity ${\mathbf V}$ and $2N$ auxiliary variables
${\mathbf U}_j$ and ${\mathbf Z}_j$, where $j=1,2,\dots,N$.
Denoting ${\mathbf X} \equiv (X_1,X_2,X_3)$,
${\mathbf V} \equiv (V_1,V_2,V_3)$, 
${\mathbf U}_j \equiv (U_{j,1},U_{j,2},U_{j,3})$
and ${\mathbf Z}_j \equiv (Z_{j,1},Z_{j,2},Z_{j,3})$, 
the time evolution of the SCG model is given by 
\begin{eqnarray}
\mbox{d}X_i & = & V_{i} \, \mbox{d}t, 
{\hskip 5.76cm}
\mbox{for}
\;\;
i=1,2,3,
\label{scm1}
\\
\mbox{d}V_i & = & \sum_{j=1}^N U_{j,i} \, \mbox{d}t, 
\label{scm2}
\\
\mbox{d}U_{j,i} & = & 
\left(
-\eta_{j,1} V_i + h_j(Z_{j,i})
\right)
\,
g_j^\prime(g_j^{-1}(U_{j,i}))
\, \mbox{d}t,
{\hskip 1cm}
\mbox{for}
\;\;
j=1,2,\dots,N, \qquad
\label{scm3}
\\
\mbox{d}Z_{j,i} \rule{0pt}{6mm}
& = & 
- 
\left( 
\eta_{j,2} \, h_j(Z_{j,i})
+
\eta_{j,3} {\hskip 0.2mm} U_{j,i} \right)
\, \mbox{d}t
+ 
\eta_{j,4} \; \mbox{d}W_{j,i} {\hskip 0.2mm} , 
\label{scm4}
\end{eqnarray}
where $g_j: {\mathbb R} \to {\mathbb R}$ is an increasing differentiable 
function, $g_j^\prime$ is its derivative,
$g_j^{-1}$ is its inverse, $h_j: {\mathbb R} \to {\mathbb R}$
is a continuous function and $\eta_{j,k}$ are positive constants
for $j=1,2,\dots,N$ and $k=1$, $2$, $3$, $4$. We note that some 
of our assumptions on $g_j$ can be relaxed as long 
as $g_j^\prime(g_j^{-1}(U_{j,i}))$
appearing in equation~(\ref{scm3}) can be suitably defined. 

The SCG description (\ref{scm1})--(\ref{scm4}) includes
$2N$ functions $g_j$ and $h_j$ and $4N$ additional parameters $\eta_{j,k}$, 
which can be all adjusted to fit properties of the detailed all-atom MD model.
In particular the SCG model (\ref{scm1})--(\ref{scm4})
can better match the MD trajectories of ions than the BD description 
given by equation~(\ref{BDSDE}), which only has one parameter, diffusion 
constant $D$, to fit to the MD results. 

One of the shortcomings of equation
(\ref{BDSDE}) is that its derivation from the underlying MD model requires 
us to consider the limit
of sufficiently large times. In particular, we need to discretize equation 
(\ref{BDSDE}) with a relatively large time step, say a nanosecond, to use it 
as a description of the trajectory of an ion. Since the typical time step of 
an all-atom MD model is a femtosecond, it is difficult 
to design a multi-resolution scheme which would replace 
all-atom MD simulations by equation (\ref{BDSDE}) in parts of 
the computational domain. The SCG model (\ref{scm1})--(\ref{scm4}) 
can be used to fit not only the diffusion constant $D$ but other important 
properties of all-atom MD models, which improves the accuracy of the 
SCG model at time steps comparable with the MD timestep. 

SCG models can be constructed using a relatively automated procedure by 
postulating that an ion interacts with additional
`fictitious particles'. Such a methodology has been applied to
coarse-grained modelling of biomolecules by~\citet{Davtyan:2015:DFM,Davtyan:2016:DFM}
to improve the fit between an MD model and the dynamics on a 
coarse-grained potential surface. They use fictitious particles with 
harmonic interactions with coarse-grained degrees of freedom 
(i.e. they add quadratic terms to the potential function 
of the system and linear terms to equations of motions) 
and each fictitious particle is also subject to a friction force and 
noise. An application of such an approach to ions leads to systems of 
linear stochastic differential equations (SDEs) and can be used, after some
transformation, to obtain a simplified version of the SCG model 
(\ref{scm1})--(\ref{scm4}), where functions $g_j$ and $h_j$ are given as identities,
i.e. $g_j(y) = h_j(y)=y$ for $y \in {\mathbb R}$ and 
$j=1,2,\dots,N$. Using this simplifying assumption in 
the SCG model~(\ref{scm1})--(\ref{scm4}), we obtain
\begin{eqnarray}
\mbox{d}X_i & = & V_{i} \, \mbox{d}t, 
{\hskip 4.5cm}
\mbox{for}
\;\;
i=1,2,3,
\label{scm1sim}
\\
\mbox{d}V_i & = & \sum_{j=1}^N U_{j,i} \, \mbox{d}t, 
\label{scm2sim}
\\
\mbox{d}U_{j,i} & = & 
\left(
-\eta_{j,1} V_i + Z_{j,i}
\right)
\, \mbox{d}t,
{\hskip 2.38cm}
\mbox{for}
\;\;
j=1,2,\dots,N, \qquad
\label{scm3sim}
\\
\mbox{d}Z_{j,i} \rule{0pt}{6mm}
& = & 
- 
\left( 
\eta_{j,2} Z_{j,i}
+
\eta_{j,3} {\hskip 0.2mm} U_{j,i} \right)
\, \mbox{d}t
+ 
\eta_{j,4} \; \mbox{d}W_{j,i} {\hskip 0.2mm} .
\label{scm4sim}
\end{eqnarray}
This is a linear system of SDEs with $4N$
parameters. It has been shown by \citet{Erban:2016:CAM} that such
models can fit an increasing number of properties of all-atom MD 
simulations as we increase $N$. For example, the linear SCG
model (\ref{scm1sim})--(\ref{scm4sim}) can be used
to fit the diffusion constant~$D$ and second moments of the velocity
and the force for $N=1$, while the velocity autocorrelation function
can better be fitted for larger values of $N$, e.g. for $N=3$. However, 
there are other properties of MD simulations which cannot be captured 
by linear models even if consider arbitrarily large $N$. They include, for 
example, all distributions which are not Gaussian. This motivates
the introduction of general functions $h_j$ and $g_j$
in the SCG model (\ref{scm1})--(\ref{scm4}).

Considering the SCG model (\ref{scm1})--(\ref{scm4})
in its full generality, it can capture more interesting dynamics. However,
coarse-grained models can only be useful if they can be easily parametrized.
Thus in our analysis, we focus on choices of functions $g_j$ and $h_j$
which both improve the properties of the SCG description
and do not complicate its analysis and parametrization.
The rest of the paper is organized as follows. In Section~\ref{secN1lin}, 
we consider the linear SCG model (\ref{scm1sim})--(\ref{scm4sim}) 
for $N=1$, which is followed in Section~\ref{secgenNlin} with the analysis
of the linear model for general values of $N$. To get some further insights
into the properties of this model, we study its connections with the 
corresponding generalized Langevin equation. In Section~\ref{secN1nonlin},
we consider the nonlinear SCG model (\ref{scm1})--(\ref{scm4}) for $N=1$.
We consider specific choices of nonlinearity $g_1$, for which the
model can be solved in terms of incomplete gamma functions. This helps
us to design three approaches to parametrize the nonlinear SCG model,
which are applied to data obtained from MD simulations. We conclude
with the analysis of the nonlinear SCG model (\ref{scm1})--(\ref{scm4}) for 
general values of $N$ in Section~\ref{secgenNnonlin}.

\section{Linear model for $N=1$ and the generalized Langevin equation}

\label{secN1lin}

We begin by considering the linear SCG model~(\ref{scm1sim})--(\ref{scm4sim}) 
for $N=1$. To simplify our notation in this section, we will drop some 
subscripts and denote $X = X_i,$ $V = V_i$, $U = U_{1,i}$, $Z=Z_{1,i}$,
$W=W_{1,i}$ and $\eta_{k} = \eta_{1,k}$ for $k=1$, $2$, $3$, $4$.
Then equations (\ref{scm1sim})--(\ref{scm4sim}) read as follows
\begin{eqnarray}
\mbox{d}X & = & V \, \mbox{d}t, 
\label{scm1simsim}
\\
\mbox{d}V & = & U \, \mbox{d}t, 
\label{scm2simsim}
\\
\mbox{d}U & = & 
\left(
-\eta_{1} V + Z
\right)
\,
\mbox{d}t,
\label{scm3simsim}
\\
\mbox{d}Z
& = & 
- 
\left( 
\eta_{2} Z
+
\eta_{3} {\hskip 0.2mm} U \right)
\, \mbox{d}t
+ 
\eta_{4} \; \mbox{d}W, 
\qquad
\label{scm4simsim}
\end{eqnarray}
where $X$ is (one coordinate of) the position of the coarse-grained particle
(ion), $V$ is its velocity, $U$ is its acceleration, 
$Z$ is an auxiliary variable, $\mbox{d}W$ is white 
noise and $\eta_j$, $j=1$, $2$, $3$, $4$, are positive parameters. In order
to find the values of four parameters $\eta_j$ suitable for modelling ions, 
\citet{Erban:2016:CAM} estimates the diffusion constants $D$ and three second moments 
$\langle V^2 \rangle$, $\langle U^2 \rangle$ and $\langle Z^2 \rangle$
from all-atom MD simulations of ions (K$^+,$ Na$^+,$
Ca$^{2+}$ and Cl$^-$) in aqueous solutions. The four parameters
of the SCG model (\ref{scm1simsim})--(\ref{scm4simsim})
can then be chosen as
\begin{equation}
\eta_1
=
\frac{\langle U^2 \rangle}{\langle V^2 \rangle},
\;\;
\eta_2
=
\frac{\langle Z^2 \rangle}{D}
\left(
\frac{\langle V^2 \rangle}{\langle U^2 \rangle}
\right)^{\!\!2},
\;\;
\eta_3
=
\frac{\langle Z^2 \rangle}{\langle U^2 \rangle},
\;\;
\eta_4
=
\sqrt{\frac{2}{D}}
\frac{\langle V^2 \rangle \langle Z^2 \rangle}{\langle U^2 \rangle}.
\label{etaestimates}
\end{equation}
Then the SCG model (\ref{scm1simsim})--(\ref{scm4simsim})
gives the same values of $D$, $\langle V^2 \rangle$, $\langle U^2 \rangle$ and 
$\langle Z^2 \rangle$ as obtained in all-atom MD simulations. 

Since the model (\ref{scm1simsim})--(\ref{scm4simsim}) only has four parameters, 
we can only hope to get the exact match of four quantities estimated from all-atom 
MD. To get some insights into what we are missing, we will derive the corresponding
generalized Langevin equation and study its consequences. The generalized
Langevin equation can be written in the form
\begin{equation}
\frac{\mbox{d}V}{\mbox{d}t} 
= 
-
\int_0^t 
K(\tau) \, V(t-\tau) \, \mbox{d}\tau
+
R(t),
\label{genlangequation}
\end{equation}
where $K: [0,\infty) \to {\mathbb R}$ is a memory kernel and random term $R(t)$ 
satisfies the generalized fluctuation-dissipation theorem, given below
in equation~(\ref{noisecorfun}). To derive the generalized Langevin equation~(\ref{genlangequation}),
consider the two-variable subsystem (\ref{scm3simsim})--(\ref{scm4simsim})
of the SCG model. Denoting ${\mathbf y} = (U,Z)^{\mathrm T},$ where 
${\mathrm T}$ stands for transpose, equations~(\ref{scm3simsim})--(\ref{scm4simsim})
can be written in vector notation as follows
\begin{equation}
\mbox{d} {\mathbf y} = B \, {\mathbf y} \, \mbox{d}t
+ {\mathbf b_1} V \, \mbox{d}t
+ {\mathbf b_2} \, \mbox{d}W,
\label{vectorVUZsystem}
\end{equation} 
where matrix $B \in {\mathbb R}^{2 \times 2}$ and vectors
${\mathbf b}_j \in {\mathbb R}^{2}$, $j=1,2,$ are given as
$$
B
=
\left(
\begin{matrix}
0 & 1 \\
- \eta_3 & -\eta_2 
\end{matrix} 
\right),
\qquad
{\mathbf b}_1
=
\left(
\begin{matrix}
- \eta_1 \\
0
\end{matrix} 
\right)
\qquad
\mbox{and}
\qquad
{\mathbf b}_2
=
\left(
\begin{matrix}
0 \\
\eta_4
\end{matrix} 
\right).
$$ 
Let us denote the eigenvalues and eigenvectors of $B$ as 
$\lambda_j$ and 
${\pmb \nu}_j = (1, \lambda_j)^{\mathrm T},$ 
$j=1,2,$ respectively.
The eigenvalues of $B$ are the solutions of the characteristic 
polynomial 
$
\lambda^2
+ \eta_2 \, \lambda 
+ \eta_3 = 0.
$
They are given by
\begin{equation}
\lambda_{1} 
= 
- \frac{\eta_2}{2} + \mu
\quad\;
\mbox{and}
\quad\;
\lambda_{2} 
= 
- \frac{\eta_2}{2} - \mu
\quad\;
\mbox{where}
\quad\;
\mu
= 
\sqrt{\frac{\eta_2^2}{4} - \eta_3}.
\label{lam12}
\end{equation}
Since $\eta_2$ and $\eta_3$ are positive parameters$,$ we conclude that real parts of 
both eigenvalues are negative. In what follows, we will assume $\eta_2^2 \ne 4 \eta_3$. 
Then we have two distinct eigenvalues and the  general solution of the SDE system 
(\ref{vectorVUZsystem}) can be written as follows
\begin{equation}
{\mathbf y}(t)
=
\Phi(t) \, {\mathbf c}
+
\Phi(t) \int_0^t \Phi^{-1}(s) \, {\mathbf b_1} V(s) \, \mbox{d}s
+
\Phi(t) \int_0^t \Phi^{-1}(s) \, {\mathbf b_2} \, \mbox{d}W,
\label{gensolutionVUZsystem}
\end{equation} 
where ${\mathbf c} \in {\mathbb R}^{2}$ is a constant vector 
determined by initial conditions and matrix
$\Phi(t) \in {\mathbb R}^{2 \times 2}$
is given as
$$
\Phi(t) 
= 
(\exp(\lambda_1 t) {\pmb \nu}_1 \; | \;
 \exp(\lambda_2 t) {\pmb \nu}_2 )
=
\left(
\begin{matrix}
\exp(\lambda_1 t) & \exp(\lambda_2 t) \\
\lambda_1 \exp(\lambda_1 t) & \lambda_2 \exp(\lambda_2 t) 
\end{matrix} 
\right),
$$
i.e. each column is a solution of the ODE system
$\mbox{d} {\mathbf y} = B \, {\mathbf y} \, \mbox{d}t$.
Calculating the inverse of $\Phi(t)$ and considering 
long-time behaviour, equation (\ref{gensolutionVUZsystem}) 
simplifies to
\begin{equation}
U(t)
=
-
\int_0^t 
K(\tau) \, V(t-\tau) \, \mbox{d}\tau
+
R(t),
\label{gensolutionVUZsystem3}
\end{equation}
where memory kernel $K(\tau)$ is given by
\begin{equation}
K(\tau)
=
\frac{\eta_1}{\lambda_1-\lambda_2}
\left(
\lambda_1 \exp(\lambda_2 \, \tau) 
-
\lambda_2 \exp(\lambda_1 \, \tau)
\right)
\label{kernelKs}
\end{equation}
and noise term 
$R(t)$ is Gaussian with zero mean and the equilibrium correlation function 
satisfying the generalized fluctuation-dissipation theorem in the form
\begin{equation}
\langle R(t_1) R(t_2) \rangle
=
\frac{\eta_4^2}{2 \eta_1 \eta_2 \eta_3} \, K(t_2-t_1).
\label{noisecorfun}
\end{equation}
Using (\ref{lam12}), memory kernel~(\ref{kernelKs}) can be rewritten as
\begin{equation}
K(\tau)
=
\eta_1
\,
\exp 
\left(
-
\frac{\eta_2 \, \tau}{2}
\right)
\,
\left(
\cosh
\left(
\mu \, \tau
\right)
+
\frac{\eta_2}{2 \mu}
\sinh
\left(
\mu \, \tau
\right)
\right),
\label{kernelgammatauOPSM}
\end{equation}
where $\mu = \sqrt{\eta_2^2/4 - \eta_3}$. We note that the auxiliary
coefficient $\mu$ is a square root of a real negative number for
$\eta_2^2 < 4 \eta_3$. However,  formula (\ref{kernelgammatauOPSM})
is still valid in this case: for $\eta_2^2 < 4 \eta_3$ it can  be
rewritten in terms of sine and cosine functions, taking into account 
that $\mu = {\mathrm i} \, |\mu|$ is pure imaginary,
$\sinh ({\mathrm i} \, |\mu| \, \tau) = {\mathrm i} \, \sin (|\mu|) \, \tau$ and
$\cosh ({\mathrm i} \, |\mu| \, \tau) = \cos (|\mu| \, \tau).$

\begin{figure}
\centerline{
\hskip 1mm
\raise 4.5cm \hbox{\raise 0.9mm \hbox{(a)}}
\hskip -6mm
\epsfig{file=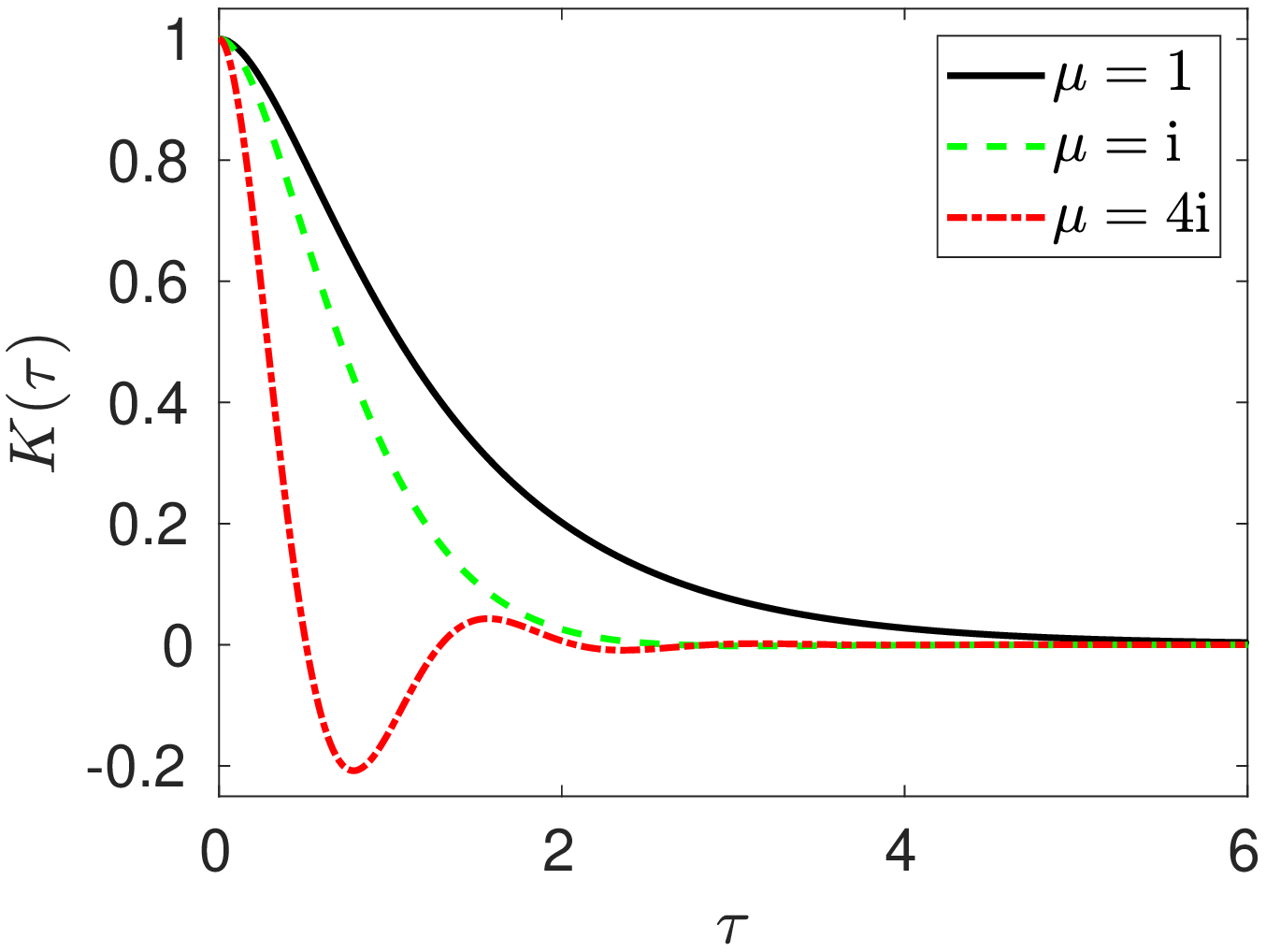,height=4.5cm}
\hskip 3mm
\raise 4.5cm \hbox{\raise 0.9mm \hbox{(b)}}
\hskip -6mm
\epsfig{file=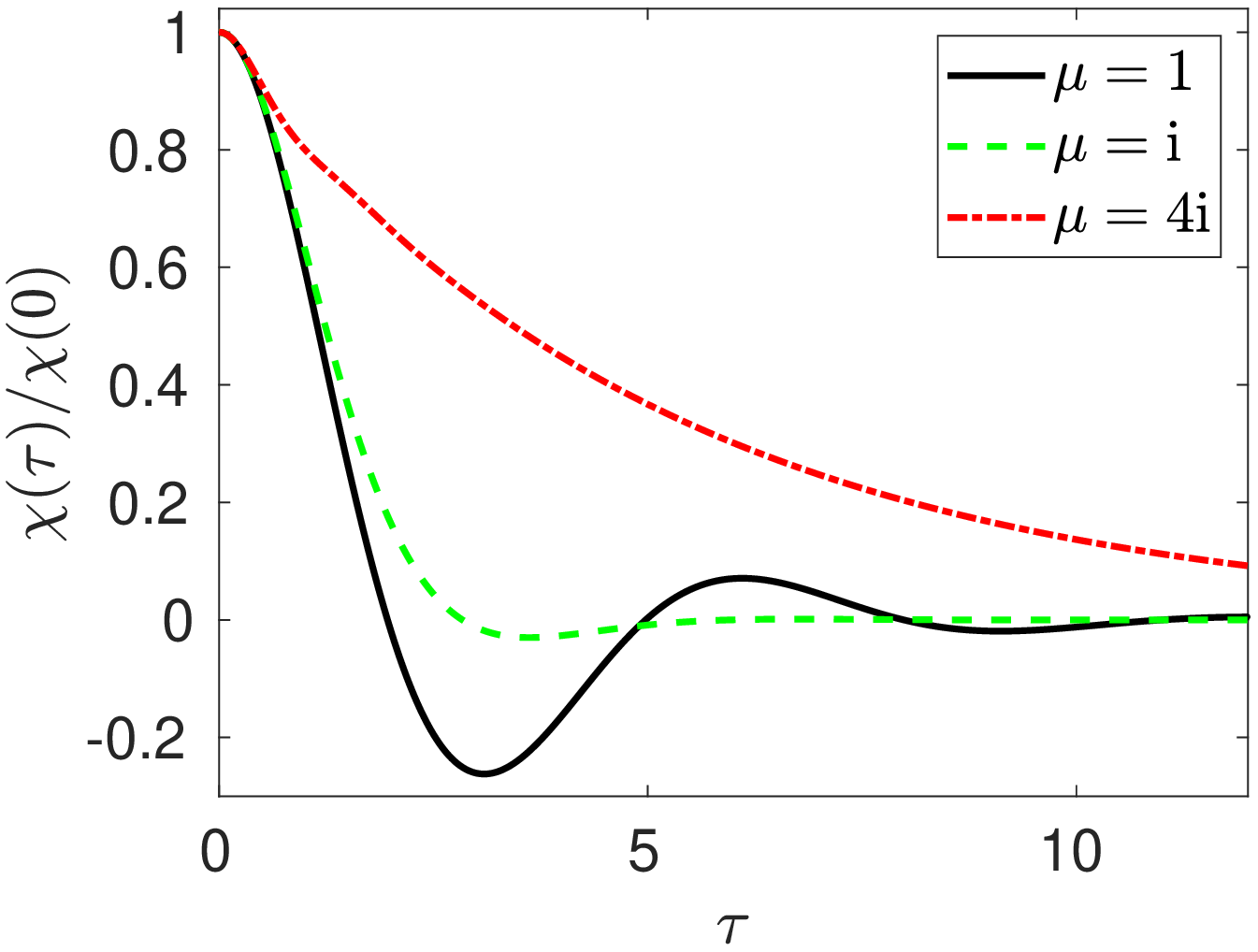,height=4.5cm}
}
\caption{(a) {\it Memory kernel $K(\tau)$ given by equation $(\ref{kernelgammatauOPSM})$
for $\eta_1 = 1$, $\eta_2 = 4$ and three different values of $\eta_3$, namely
$\eta_3 = 3$ (solid line, $\mu=1$), $\eta_3 = 5$ (dashed line, $\mu={\mathrm i}$) 
and $\eta_3 = 20$ (dot-dashed line, $\mu=4 {\mathrm i}$).}
(b) {\it Normalized velocity autocorrelation function $\chi(\tau)/\chi(0)$
computed by using equation $(\ref{chilaplaceOPSM2})$ for the same parameter values
as in panel} (a).}
\label{figure1} 
\end{figure}

The memory kernel $K(\tau)$, given by equation (\ref{kernelgammatauOPSM}), 
is plotted in Figure~\ref{figure1}(a) for different values
of parameter $\mu$. For simplicity, we use non-dimensionalized versions
of our equations with dimensionless parameters $\eta_1 = 1$ and $\eta_2 = 4$. 
We choose three different values of $\eta_3$ so that the values of $\mu$ are
1, ${\mathrm i}$ and $4 {\mathrm i}$.
In Figure~\ref{figure1}(b), we plot the equilibrium velocity 
autocorrelation function which is defined as
$$
\chi(\tau)
=
\lim_{t \to \infty}
\langle V(t) \, V(t-\tau) \rangle,
$$
for $\tau \in [0,\infty)$. More precisely, we plot $\chi(\tau)/\chi(0)$ which is
normalized so that its value at $\tau=0$ is equal to 1. It is related
to the memory kernel by
\begin{equation}
\frac{\chi(\tau)}{\chi(0)}
=
\mathscr{L}^{-1} \!
\left( \frac{1}{s + \mathscr{L}\big[K\big](s)} \right) \, ,
\label{chilaplace}
\end{equation}
where $\mathscr{L}\big[K\big](s) = \int_0^\infty
K(\tau) \exp(-s \tau) \, \dtau$ is the Laplace transform of the memory kernel
$K(\tau)$ and $\mathscr{L}^{-1}$ denotes Laplace inversion. 
Following~\citet{Erban:2019:SMR}, we evaluate the right hand side of equation
(\ref{chilaplace}) as follows. Substituting equation 
(\ref{kernelgammatauOPSM}) into (\ref{chilaplace}), we obtain
\begin{equation}
\frac{\chi(\tau)}{\chi(0)}
=
\mathscr{L}^{-1} \!
\left( \frac{s^2 + \eta_2 s + \eta_3}
{s^3 + \eta_2 s^2 + (\eta_1 + \eta_3) s + \eta_1 \eta_2}
\right) \, .
\label{chilaplace2}
\end{equation}
The polynomial in the denominator, 
$p(s)=s^3 + \eta_2 s^2 + (\eta_1 + \eta_3) s 
+ \eta_1 \eta_2,$
has positive coefficients. Since $p(-\eta_2) < 0 < p(0)$, 
it has one negative
real root in interval $(-\eta_2,0)$, which we denote 
by $a_1$. The other two roots ($a_2$ and $a_3$ say) may be real or
complex, but if they are complex they will be complex conjugates since
$p(s)$ has real coefficients. Assuming that the real part of each
root is negative, we first find the partial fraction decomposition
of the rational function in (\ref{chilaplace2}) as
$$
\frac{s^2 + \eta_2 s + \eta_3}
{s^3 + \eta_2 s^2 + (\eta_1 + \eta_3) s + \eta_1 \eta_2}
=
\frac{c_1}{s-a_1}
+
\frac{c_2}{s-a_2}
+
\frac{c_3}{s-a_3},
$$
where $c_i \in {\mathbb C}$ are constants (which depend on $\eta_1$,
$\eta_2$ and $\eta_3$). Then we can rewrite (\ref{chilaplace})
as
\begin{equation}
\frac{\chi(\tau)}{\chi(0)}
=
c_1 \exp(a_1 \tau)
+
c_2 \exp(a_2 \tau)
+
c_3 \exp(a_3 \tau)\, .
\label{chilaplaceOPSM2}
\end{equation}
The results computed by (\ref{chilaplaceOPSM2}) are shown in Figure~\ref{figure1}(b). 
We note that although equation (\ref{chilaplaceOPSM2}) may include complex  exponentials,
the resulting $\chi(\tau)$ is always real. Since the diffusion constant, $D$, and
the second moment of the equilibrium velocity distribution, 
$\langle V^2 \rangle$, are related to $\chi$ by
$$
D
=
\int_0^\infty \!\! \chi(\tau) \, \dtau
=
\frac{\eta_4^2}{2 \, \eta_1^2 \, \eta_2^2} 
\qquad
\mbox{and}
\qquad
\langle V^2 \rangle
=
\chi(0)
=
\frac{\eta_4^2}{2 \, \eta_1 \, \eta_2 \, \eta_3} \, ,
$$
the parametrization (\ref{etaestimates}) guarantees that both the value
of $\chi(0)$ and the integral of $\chi(\tau)$ are captured accurately. However,
the simplified SCG description (\ref{scm1simsim})--(\ref{scm4simsim})
is not suitable to perfectly fit the velocity autocorrelation function
or the memory kernel for all values of $\tau \in [0,\infty)$. 
In order to do this, we have to consider the
SCG model (\ref{scm1sim})--(\ref{scm4sim})
for larger values of $N$ as it is done in the following section.

\section{General linear SCG model and autocorrelation functions}

\label{secgenNlin}

Considering the linear SCG model~(\ref{scm1sim})--(\ref{scm4sim}) for general
values of $N$, we can solve equations (\ref{scm3sim})--(\ref{scm4sim}) for
each value of $j=1$, $2$, $\dots$, $N$ to generalize our previous
result (\ref{gensolutionVUZsystem3}) as
\begin{equation}
U_{j,i}(t)
=
-
\int_0^t 
K_j(\tau) \, V_i(t-\tau) \, \mbox{d}\tau
+
R_{j,i}(t) \, ,
\label{gensolutionUji}
\end{equation}
where kernel $K_j(\tau)$ is given by (compare with (\ref{kernelgammatauOPSM}))
\begin{equation}
K_j(\tau)
=
\eta_{j,1}
\,
\exp 
\left(
-
\frac{\eta_{j,2} \, \tau}{2}
\right)
\,
\left(
\cosh
\left(
\mu_j \, \tau
\right)
+
\frac{\eta_{j,2}}{2 \mu_j}
\sinh
\left(
\mu_j \, \tau
\right)
\right)
\label{kernelKjs}
\end{equation}
with
\begin{equation}
\mu_j = \sqrt{\frac{\eta_{j,2}^2}{4} - \eta_{j,3}}
\label{parmuj}
\end{equation} 
and noise term $R_{j,i}(t)$ is Gaussian with zero mean and the equilibrium correlation function 
satisfying
$$
\langle R_{j,i}(t_1) R_{j,i}(t_2) \rangle
=
\frac{\eta_{j,4}^2}{2 \, \eta_{j,1} \, \eta_{j,2} \, \eta_{j,3}} \, K_j(t_2-t_1).
$$
Substituting (\ref{gensolutionUji}) to (\ref{scm2sim}), we obtain the generalized
Langevin equation
\begin{equation}
\frac{\mbox{d}V_i}{\mbox{d}t} 
= 
-
\int_0^t 
K(\tau) \, V_i(t-\tau) \, \mbox{d}\tau
+
R_i(t) \, ,
\label{genlangequation2}
\end{equation}
where 
\begin{equation}
K(\tau)
=
\sum_{j=1}^N K_j(\tau)
\qquad \mbox{and} \qquad
R_i(t)
=
\sum_{j=1}^N R_{j,i}(t).
\label{KRequations}
\end{equation}
In particular, we have $3N$ parameters to fit memory kernel $K(\tau)$, which can
be estimated from all-atom MD simulations. There have been a number of approaches
developed in the literature to estimate the memory kernel from MD simulations.
\citet{Shin:2010:BMM} use an integral equation with relates memory kernel $K(\tau)$
with the autocorrelation function for the force and the correlation function between 
the force and the velocity. Estimating these correlation functions from 
long time MD simulations and solving the integral equation, they obtain 
memory kernel $K(\tau)$. Other methods to estimate the memory kernel, $K(\tau)$, 
of the corresponding generalized Langevin equation~(\ref{genlangequation2}) 
have been presented by~\citet{Gottwald:2015:PLG} and \citet{Jung:2017:IRM}.

An alternative approach to parametrize the linear SCG model~(\ref{scm1sim})--(\ref{scm4sim})
is to estimate the velocity autocorrelation function, $\chi(\tau)$, from all-atom MD 
simulations. This can be done by computing how correlated is the current velocity 
(at time $t$) with velocity at previous times. Since equations 
(\ref{scm1simsim})--(\ref{scm4simsim}) are linear SDEs, 
we can follow \citet{Mao:2007:SDE} to solve them analytically, using eigenvalues
and eigenvectors of matrices appearing in their corresponding matrix formulation.
Using this analytic solution, \citet{Erban:2016:CAM} use an acceptance-rejection
algorithm to fit the parameters of linear SCG model~(\ref{scm1sim})--(\ref{scm4sim}) 
for $N=3$ to match the velocity autocorrelation functions of ions estimated from
all-atom MD simulations of Na$^+$ and K$^+$ in the SPC/E water.

Since the parameter $\mu_j$ given by (\ref{parmuj}) is a square root of a real number,
it can be both positive or purely imaginary. In particular, kernels $K_j(\tau)$ 
given by equation~(\ref{kernelKjs}) can include both exponential, sine and cosine functions
as illustrated in Figure~\ref{figure1}(a). Since memory kernel $K(\tau)$ is given
as the sum of $K_j(\tau)$ in equation (\ref{KRequations}), typical memory kernels 
and correlation functions estimated from all-atom MD simulations can be 
successfully matched by linear SCG models for relatively small values of $N$. However, as shown by
\citet{Mao:2007:SDE}, analytic solutions of linear SDEs also imply that the process 
is Gaussian at any time $t>0$, provided that we start with deterministic initial 
conditions. Thus the linear SCG model~(\ref{scm1sim})--(\ref{scm4sim}) for
abtitrary values of $N$ can only fit distributions which are Gaussian. This motivates
our investigation of the nonlinear SCG model in the next two sections.

\section{Nonlinear SCG model for $N=1$}

\label{secN1nonlin}

We begin by considering the nonlinear SCG model~(\ref{scm1})--(\ref{scm4}) 
for $N=1$. As in Section~\ref{secN1lin}, we simplify our 
notation by dropping some subscripts and denoting $X = X_i,$ $V = V_i$, 
$U = U_{1,i}$, $Z=Z_{1,i}$, $W=W_{1,i}$, $g=g_j$, $h=h_j$ 
and $\eta_{k} = \eta_{1,k}$ for $k=1$, $2$, $3$, $4$. 
Then equations (\ref{scm1})--(\ref{scm4}) read as follows
\begin{eqnarray}
\mbox{d}X & = & V \, \mbox{d}t, 
\label{scm1nonlin}
\\
\mbox{d}V & = & U \, \mbox{d}t, 
\label{scm2nonlin}
\\
\mbox{d}U & = & 
\left(
-\eta_{1} V + h(Z)
\right)
\,
g^\prime(g^{-1}(U))
\, 
\mbox{d}t,
\label{scm3nonlin}
\\
\mbox{d}Z
& = & 
- 
\left( 
\eta_{2} \, h(Z)
+
\eta_{3} \, U \right)
\, \mbox{d}t
+ 
\eta_{4} \; \mbox{d}W, 
\qquad
\label{scm4nonlin}
\end{eqnarray}
where $X$ denotes (one coordinate of) the position of the coarse-grained particle, 
$V$ is its velocity, $U$ is its acceleration, 
$Z$ is an auxiliary variable, $\mbox{d}W$ is white 
noise, $\eta_j$, for $j=1$, $2$, $3$, $4$, are positive parameters and
functions $g: {\mathbb R} \to {\mathbb R}$ 
and $h: {\mathbb R} \to {\mathbb R}$ are yet to be specified.

Equation~(\ref{scm1nonlin}) describes the time evolution of the 
position, while equations (\ref{scm2nonlin})--(\ref{scm4nonlin}) admit 
a stationary distribution. We denote it by $p(v,u,z)$.
Then $p(v,u,z) \, \mbox{d}v \, \mbox{d}u \, \mbox{d}z$ 
gives the probability that $V(t) \in [v,v+\mbox{d}v)$, $U(t) \in [u,u+\mbox{d}u)$ 
and $Z(t) \in [z,z+\mbox{d}z)$ at equilibrium.
The stationary distribution, $p(v,u,z)$, of SDEs (\ref{scm2nonlin})--(\ref{scm4nonlin}) 
can be obtained by solving the corresponding stationary Fokker-Planck equation
\begin{eqnarray*}
\frac{\eta_4^2}{2}
\frac{\partial^2 p}{\partial^2 z}
(v,u,z)
&=&
\frac{\partial}{\partial v}
\Big(
u \, p(v,u,z)
\Big)
+
\frac{\partial}{\partial u}
\Big(
\big(
-
\eta_1 v
+
h(z)
\big)
g^\prime(g^{-1}(u)) \,
p(v,u,z)
\Big) \\
&& +
\frac{\partial}{\partial z}
\Big(
\big(
- \eta_2 h(z) 
- \eta_3 u
\big)
p(v,u,z)
\Big),
\end{eqnarray*}
which gives
\begin{equation}
p(v,u,z)
=
\frac{C}{g^\prime(g^{-1}(u))}
\,
\exp
\!
\left[
-
\frac{2 \eta_2}{\eta_4^2} 
\left(
\eta_1 \eta_3
\, \frac{v^2}{2}
+
\eta_3 \, G \big( g^{-1}(u) \big)
+
H(z) \right)
\right],
\label{stdisN1nonlin}
\end{equation}
where $C$ is the normalization constant, and functions $G$ and $H$ are
integrals of functions $g$ and $h$, respectively, which are given by
\begin{equation}
G(y) = \int_0^y g(\xi) \, \dxi
\qquad
\mbox{and}
\qquad
H(y) = \int_0^y h(\xi) \, \dxi.
\label{GHdefin}
\end{equation}
We note that for the special case where $g$ and $h$ are given as identities,
i.e. $g(y) = h(y)=y$ for $y \in {\mathbb R}$, the nonlinear 
SCG model~(\ref{scm1nonlin})--(\ref{scm4nonlin}) is equal to the linear 
SCG model~(\ref{scm1simsim})--(\ref{scm4simsim}) and
functions $G$ and $H$ are $G(y)=H(y)=y^2/2$. Then the stationary distribution
(\ref{stdisN1nonlin}) is product of Gaussian distributions in $v$, $u$ and
$z$ variables. In particular, we can easily calculate the second moments
of these distributions in terms of parameters $\eta_j$. Estimating these
moments from all-atom MD simulations, we can parametrize the 
resulting linear SCG model (\ref{scm1simsim})--(\ref{scm4simsim})
as shown in equation (\ref{etaestimates}). However, if we want to match
a non-Gaussian force distribution, we have to consider nonlinear models.
A simple one-parameter example is studied in the next section.

\subsection{One-parameter nonlinear function}

\label{oneparam}

\begin{figure}
\centerline{
\hskip 1mm
\raise 4.5cm \hbox{\raise 0.9mm \hbox{(a)}}
\hskip -6mm
\epsfig{file=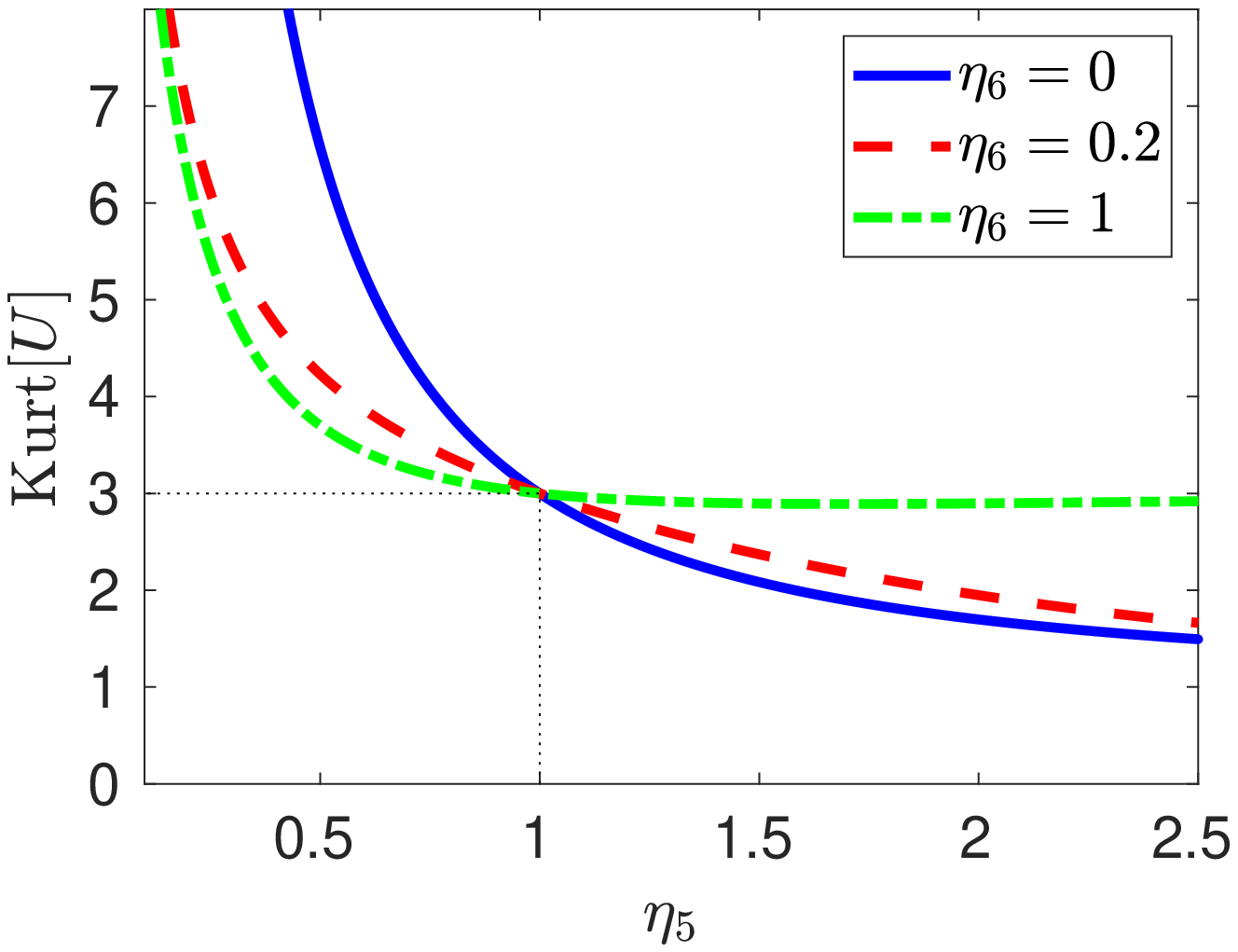,height=4.5cm}
\hskip 3mm
\raise 4.5cm \hbox{\raise 0.9mm \hbox{(b)}}
\hskip -6mm
\epsfig{file=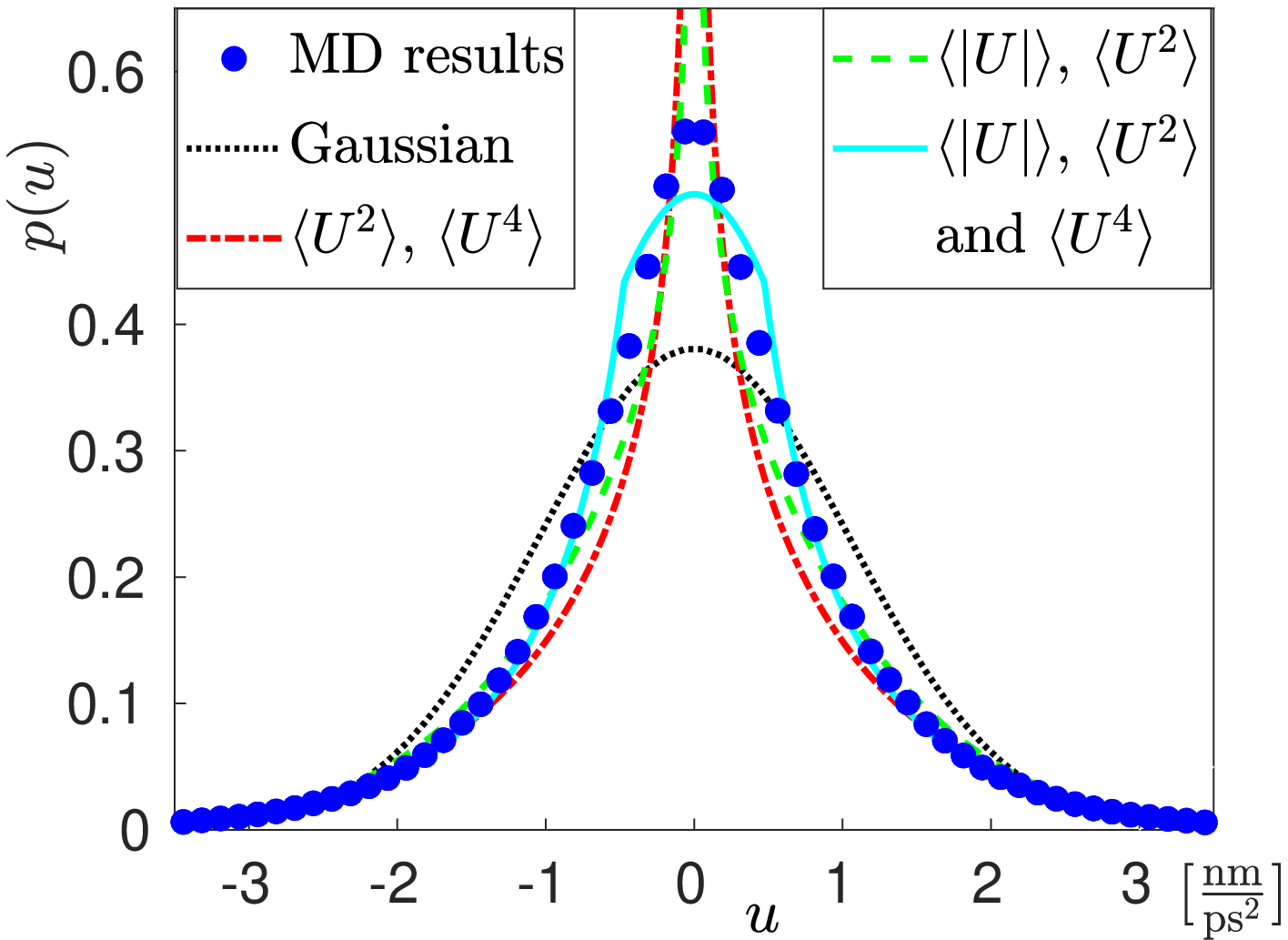,height=4.5cm}
}
\caption{(a) {\it Kurtosis $\mathrm{Kurt}[U]$ given by equation $(\ref{kurtosistwopar})$
as a function of parameter $\eta_5$ for three different values of parameter $\eta_6$. 
The result for $\eta_6=0$ (blue solid line) corresponds to the case of one-parameter function
$g$, defined by $(\ref{gdefsim})$, where the kurtosis is given by~$(\ref{kurtosis})$.
} \hfill\break
(b){\it Distribution of $U$ estimated from a long-time MD simulation (blue circles) 
compared with the results obtained 
by the linear SCG model~$(\ref{scm1simsim})$--$(\ref{scm4simsim})$ (black dotted line),
nonlinear SCG models~$(\ref{scm1nonlin})$--$(\ref{scm4nonlin})$ with 
one-parameter function $g$, defined by $(\ref{gdefsim})$, 
fitting 
$\langle U^2 \rangle$ and $\langle U^4 \rangle$ (red dot-dashed line)
and 
$\langle |U| \rangle$ and $\langle U^2 \rangle$ (green dashed line),
and the nonlinear SCG model~$(\ref{scm1nonlin})$--$(\ref{scm4nonlin})$ with
two-parameter function $g$ defined by $(\ref{gdef2an578qw})$, 
matching all three moments $\langle |U| \rangle$,
$\langle U^2 \rangle$ and $\langle U^4 \rangle$ (cyan solid line).
}}
\label{figure2} 
\end{figure}

Consider that $g$ is a function depending on one additional positive
parameter $\eta_5$ as follows
\begin{equation}
g(y) = |y|^{1/\eta_5} \sign y,
\label{gdefsim}
\end{equation}
where we use $\sign$ to denote the sign (signum) function
\begin{equation}
\sign y
=
\left\{
\begin{array}{rl}
- 1, & \qquad \mbox{for} \; y < 0, \\
0,  & \qquad \mbox{for} \; y =0, \\
1, & \qquad \mbox{for} \; y > 0. \\
\end{array}
\right.
\label{defsign}
\end{equation}
The function defined by (\ref{gdefsim}) only satisfies our assumptions on $g$ 
for $\eta_5 \in (0,1]$ as it is not differentiable at $y=0$ for $\eta_5>1$, but 
we will proceed with our analysis for any positive $\eta_5>0$. Consider that 
function $h$ is an identity, i.e. $h(y)=y$ for $y \in {\mathbb R}$, then
equations (\ref{scm1nonlin})--(\ref{scm4nonlin}) reduce to
\begin{eqnarray}
\mbox{d}X & = & V \, \mbox{d}t, 
\label{scm1nonlins}
\\
\mbox{d}V & = & U \, \mbox{d}t, 
\label{scm2nonlins}
\\
\mbox{d}U & = & 
\left(
-\eta_{1} V + Z
\right)
\,
\eta_5^{-1}
\,
|U|^{1- \eta_5}
\, 
\mbox{d}t,
\label{scm3nonlins}
\\
\mbox{d}Z
& = & 
- 
\left( 
\eta_{2} \, Z
+
\eta_{3} \, U \right)
\, \mbox{d}t
+ 
\eta_{4} \; \mbox{d}W, 
\qquad
\label{scm4nonlins}
\end{eqnarray}
where we would have to be careful, if we used this model to numerically simulate 
trajectories for $\eta_5>1$, because of possible division by zero for $U=0$
in equation (\ref{scm3nonlins}). If $\eta_5 \in (0,1]$, then we do not have
such technical issues. Using equation~(\ref{stdisN1nonlin}),
the stationary distribution is equal to 
\begin{equation}
p(v,u,z)
=
C
|u|^{\eta_5-1}
\,
\exp
\left[
-
\frac{\eta_2}{\eta_4^2} 
\left(
\eta_1 \eta_3
\, v^2
+
\frac{2 \eta_3 \eta_5}{1+\eta_5} 
|u|^{1+\eta_5}
+
\, z^2
\right)
\right],
\label{stdisN1nonlin2}
\end{equation}
where the normalization constant is given by
$$
\int_{-\infty}^\infty
\int_{-\infty}^\infty
\int_{-\infty}^\infty
p(v,u,z) \, \dv \, \du \, \dz
=
1.
$$
Integrating (\ref{stdisN1nonlin2}), we get
\begin{equation*}
C
=
\frac{\eta_2 \sqrt{\eta_1\eta_3}}{\pi \eta_4^2}  
\left( \frac{\eta_2 \eta_3 \eta_5}{\eta_4^2} \right)^{\eta_5/(1+\eta_5)}
\!\!\left( \frac{1+\eta_5}{2} \right)^{1/(1+\eta_5)}
\!\!\frac{1}{\mathrm{\Gamma} \! \left( \frac{\eta_5}{1+\eta_5} \right)} \;,  
\label{normconstgen}
\end{equation*}
where $\mathrm{\Gamma}$ is the gamma function defined as
\begin{equation}
\mathrm{\Gamma}(s) = \int_0^\infty \xi^{s-1} \exp(-\xi) \, \mbox{d} \xi.
\label{gammadef}
\end{equation}
Let $\alpha \ge 0$. Integrating (\ref{stdisN1nonlin2}), we get
\begin{equation}
\langle |U|^\alpha \rangle
=
\left(
\frac{\eta_4^2 \left( 1+\eta_5 \right)}{2 \eta_2 \eta_3 \eta_5}
\right)^{\alpha/(1+\eta_5)}
\frac{\mathrm{\Gamma} \!
\left(
\frac{\alpha+\eta_5}{1+\eta_5}
\right)
}
{\mathrm{\Gamma} \! \left( \frac{\eta_5}{1+\eta_5} \right)} \, .
\label{Ualphmomsim}
\end{equation}
Using (\ref{Ualphmomsim}) for $\alpha=2$ and $\alpha=4$, we obtain the following
expression for kurtosis
\begin{equation}
\mathrm{Kurt}[U]
=
\frac{\langle U^4 \rangle}{\langle U^2 \rangle^2}
=
\mathrm{\Gamma} \! 
\left( \frac{\eta_5}{1+\eta_5} \right)
\mathrm{\Gamma} \!
\left(
\frac{4+\eta_5}{1+\eta_5}
\right)
\left(
\mathrm{\Gamma} \!
\left(
\frac{2+\eta_5}{1+\eta_5}
\right)
\right)^{\!\!-2}.
\label{kurtosis}
\end{equation}
In particular, the kurtosis is only a function of one parameter, $\eta_5$. It is plotted in
Figure~\ref{figure2}(a) as the blue solid line, together with the kurtosis obtained for a more
general two-parameter SCG model studied in Section~\ref{twoparam}.
We observe that the distribution of $U$ is leptokurtic
for $\eta_5<1$ and platykurtic for $\eta_5>1$. If $\eta_5$ is equal to 1, then 
our SCG model given by equations~(\ref{scm1nonlin})--(\ref{scm4nonlin}) reduces to 
the linear SCG model given by equations~(\ref{scm1simsim})--(\ref{scm4simsim}),
i.e. the stationary distribution is Gaussian and its kurtosis is 3. This is 
shown by the dotted line in Figure~\ref{figure2}(a).

Since equation~(\ref{kurtosis}) only depends on parameter $\eta_5$,
we can use the kurtosis of the acceleration distribution
(which is equal to the kurtosis of the force distribution) esimated 
from MD simulations to find the value of parameter $\eta_5.$ To 
calculate the kurtosis, we estimate the fourth moment
$\langle U^4 \rangle$ in addition to the second moment,
$\langle U^2 \rangle$, used before in our estimating proceduce
(\ref{etaestimates}) for the linear model. In particular, we not
only get equation~(\ref{kurtosis}) for calculating the value
of parameter $\eta_5$, but also a restriction on
other parameters $\eta_2$, $\eta_3$ and $\eta_4$. Using
(\ref{Ualphmomsim}) for $\alpha=2$, it can be stated as follows
\begin{equation}
{\hskip -4mm}
\frac{\eta_4^2}{2 \, \eta_2 \, \eta_3}
=
\frac{\eta_5}{1+\eta_5}
\left(
\frac{1+\eta_5}{\pi}
\, 
\sin \left( \frac{\pi}{1+\eta_5}  \right) 
\langle U^2 \rangle \!
\right)^{\!\! (1+\eta_5)/2}
\!\!
\left(\!
\mathrm{\Gamma} \! \left( \frac{\eta_5}{1+\eta_5} \right)
\!\right)^{\!\! 1+\eta_5}
\! ,
\label{eta5restriction}
\end{equation}
where we have used properties of the gamma function, including
$\mathrm{\Gamma}(1+y)=y \, \mathrm{\Gamma}(y)$ and Euler's reflection formula,
$\mathrm{\Gamma}(1-y) \mathrm{\Gamma}(y) \sin(\pi y) = \pi$, to simplify the
right hand side. We note that in the Gaussian case, $\eta_5=1$, 
the right hand side of equation~(\ref{eta5restriction}) further
simplifies to
\begin{equation}
\frac{\eta_4^2}{2 \, \eta_2 \, \eta_3}
= 
\langle U^2 \rangle,
\label{gaussianrestriction}
\end{equation}
which is indeed the formula for the second moment of $U$ given by the
linear SCG model~(\ref{scm1simsim})--(\ref{scm4simsim}). Equation
(\ref{eta5restriction}) provides one restriction on four 
remaining parameters, $\eta_1$, $\eta_2$ $\eta_3$ and $\eta_4$, which 
need to be specified. This can be done by estimating three additional 
statistics from MD simulations, as in the case of the linear 
SCG model~(\ref{scm1simsim})--(\ref{scm4simsim}) 
in equation (\ref{etaestimates}). Indeed, the stationary distributions 
of $V$ and $Z$ are Gaussian with mean zero. Their second moments and 
the diffusion constant, $D$, for the nonlinear SCG model~(\ref{scm1nonlin})--(\ref{scm4nonlin})
can be calculuted as
\begin{equation}
D
=
\frac{\eta_4^2}{2 \, \eta_1^2 \, \eta_2^2} \, ,
\qquad
\langle V^2 \rangle
=
\frac{\eta_4^2}{2 \, \eta_1 \, \eta_2 \, \eta_3}
\qquad
\mbox{and}
\qquad
\langle Z^2 \rangle
=
\frac{\eta_4^2}{2 \, \eta_2}.
\label{VZsecmom}
\end{equation}
Therefore, assuming that $D$, $\langle V^2 \rangle$, $\langle Z^2 \rangle$
are obtained from MD simulations and $\eta_4^2/(2 \eta_2 \eta_3)$ is given
by (\ref{eta5restriction}), we can calculate parameters $\eta_k$ by
\begin{eqnarray}
\displaystyle
\eta_1
&=
\displaystyle
\frac{1}{\langle V^2 \rangle}
\left(
\frac{\eta_4^2}{2 \, \eta_2 \, \eta_3}
\right) \, ,
{\hskip 1cm}
\eta_2
&=
\displaystyle
\frac{\langle Z^2 \rangle \, \langle V^2 \rangle^2}{D}
\left(
\frac{\eta_4^2}{2 \, \eta_2 \, \eta_3}
\right)^{\! -2}
\, ,
\label{eta12formnl}
\\
\displaystyle
\eta_3
&=
\displaystyle
\langle Z^2 \rangle
\left( \frac{\eta_4^2}{2 \, \eta_2 \, \eta_3} \right)^{\! -1} \, ,
{\hskip 1cm}
\eta_4
&=
\displaystyle
\sqrt{\frac{2}{D}}
\,
\langle Z^2 \rangle 
\, 
\langle V^2 \rangle
\left(
\frac{\eta_4^2}{2 \, \eta_2 \, \eta_3}
\right)^{\! -1} \, .
\label{eta34formnl}
\end{eqnarray}
We note that in the Gaussian case, $\eta_5=1$, we can substitute 
equation~(\ref{gaussianrestriction}) for $\eta_4^2/(2 \eta_2 \eta_3)$
and the parametrization approach (\ref{eta12formnl})--(\ref{eta34formnl})
simplifies to equation~(\ref{etaestimates}) used in the case 
of the linear SCG model~(\ref{scm1simsim})--(\ref{scm4simsim}).
In the next subsection, we generalize formula (\ref{gdefsim}) 
to a two-parameter function and show that the parametrization approach
(\ref{eta12formnl})--(\ref{eta34formnl}) is still applicable
to the case of more general SCG models.

\subsection{Two-parameter nonlinear function}

\label{twoparam}

Consider that $g$ is a function depending on two positive 
parameters $\eta_5$ and $\eta_6$ as follows 
\begin{equation}
g(y) 
= 
\left\{
\begin{array}{ll}
0 \, , & \quad 
\displaystyle
\mbox{for} \; |y| \le 
\eta_6^{\eta_5} (1-\eta_5) \, ,
\\
\displaystyle
\left(
\eta_6 \left(1 - \frac{1}{\eta_5}\right)
+ 
\frac{\eta_6^{1-\eta_5}}{\eta_5}
|y|
\right)
\sign y \, ,
& \quad 
\displaystyle
\mbox{for} \; 
\eta_6^{\eta_5} (1-\eta_5)
< |y| \le 
\eta_6^{\eta_5} \, , \\
\displaystyle
\left|
y
\right|^{1/\eta_5}
\sign y \, , 
& \quad 
\displaystyle
\mbox{for} \; |y| > 
\eta_6^{\eta_5} \, ,
\\
\end{array}
\right.
\label{gdef2an578qw}
\end{equation}
where $\sign$ function is defined by (\ref{defsign}).
In particular, our expression for function $g$ is equal to the formula
(\ref{gdefsim}) for sufficiently large values of $|y|$. As discussed
in the previous section, if we used formula (\ref{gdefsim}), there
would be some issues for $y$ close to zero (for example, the division by zero 
for $U=0$ and $\eta_5>1$ in equation (\ref{scm3nonlins})),
so our generalized formula (\ref{gdef2an578qw}) replaces (\ref{gdefsim}) 
with a linear function for smaller values of $|y|$. On the face of it, it looks
that there could also be some issues with the generalized formula (\ref{gdef2an578qw}), 
because it is not strictly increasing for $|y| \le \eta_6^{\eta_5} (1-\eta_5)$. 
However, function (\ref{gdef2an578qw}) is increasing and invertible away of this 
region with its inverse given by
$$
g^{-1}(u) 
= 
\left\{
\begin{array}{ll}
\displaystyle
\eta_5 \eta_6^{\eta_5-1}
\left(
|u|
-
\eta_6 \left(1 - \frac{1}{\eta_5}\right)
\right)
\sign u,
& \quad 
\displaystyle
\mbox{for} \; 
0 < |u| \le \eta_6, \\
\displaystyle
\left|
u
\right|^{\eta_5}
\sign u, 
& \quad 
\displaystyle
\mbox{for} \; |u| > \eta_6. 
\\
\end{array}
\right.
$$
Moreover, what we really need in equations (\ref{scm1nonlin})--(\ref{scm4nonlin}) is 
$g^\prime(g^{-1}(u))$ which can be defined as the following continuous function
\begin{equation}
g^\prime(g^{-1}(u)) 
= 
\frac{1}{\eta_5}
\times
\left\{
\begin{array}{ll}
\displaystyle
\eta_6^{1-\eta_5} \, ,
& \quad 
\displaystyle
\mbox{for} \; 
|u| \le \eta_6, \\
\displaystyle
\left|
u
\right|^{1-\eta_5} \, ,
\rule{0pt}{5.5mm} 
& \quad 
\displaystyle
\mbox{for} \; |u| > \eta_6,
\\
\end{array}
\right.
\label{gprimeinv2an578qw}
\end{equation}
where the removable discontinuity at $u=0$ has disappeared because we have defined 
$g^\prime(g^{-1}(0))=\eta_6^{1-\eta_5}/\eta_5$. 
Integrating (\ref{gdef2an578qw}) and substituting (\ref{gprimeinv2an578qw}),
we get 
\begin{equation}
G\big(g^{-1}(u)\big)
= 
\left\{
\begin{array}{ll}
\displaystyle 
\frac{\eta_5\eta_6^{\eta_5-1}}{2}
u^2,
& \quad 
\displaystyle
\mbox{for} \; 
|u| \le \eta_6, \\
\displaystyle
\frac{\eta_5(\eta_5-1) \eta_6^{1+\eta_5}}{2 (1+\eta_5)}
+
\frac{\eta_5}{1+\eta_5} 
|u|^{1+\eta_5}, 
& \quad 
\displaystyle
\mbox{for} \; |u| > \eta_6,
\\
\end{array}
\right.
\label{gintuan578qw}
\end{equation}
where $G$ is the integral of function $g$ defined by (\ref{GHdefin}).
Consider again that $h$ is an identity, i.e. $h(y)=y$ for $y \in {\mathbb R}$.
Then the stationary distribution (\ref{stdisN1nonlin}) is again Gaussian in $V$ 
and $Z$ variables with their second moments given by equation~(\ref{VZsecmom}). Let us
denote the marginal stationary distribution of $U$ by
$$
p_u(u)
=
\int_{-\infty}^\infty
\int_{-\infty}^\infty
p(v,u,z) \, \dv \, \dz.
$$
Using (\ref{stdisN1nonlin}) and (\ref{gintuan578qw}), we have
\begin{equation}
p_u(u)
=
\left\{
\begin{array}{ll}
\displaystyle
C_u
\,
\eta_6^{\eta_5-1}
\,
\exp
\left[
- 
\frac{\eta_2 \eta_3 \eta_5 \eta_6^{1+\eta_5}}{\eta_4^2}
\left(
\frac{u^2}{\eta_6^{2}}
+
\frac{1-\eta_5}{1+\eta_5}
\right)
\right],
& \quad \mbox{for} \; |u| \le \eta_6, \\
\displaystyle
C_u
\,
|u|^{\eta_5-1}
\,
\exp
\left[
- 
\frac{2 \eta_2 \eta_3 \eta_5}{\eta_4^2 (1+\eta_5)} 
|u|^{1+\eta_5}
\right],
& \quad \mbox{for} \; |u| > \eta_6, \\
\end{array}
\right.
\label{stdismarguqw}
\end{equation}
where $C_u$ is the normalization constant given by
\begin{equation*}
\int_{-\infty}^\infty
p_u(u) \, \du
=
1.
\end{equation*}
Let us define 
\begin{equation}
\kappa_1 = \frac{\eta_2 \eta_3 \eta_5 \eta_6^{1+\eta_5}}{\eta_4^2}
\qquad 
\mbox{and} 
\qquad
\kappa_2 = \frac{1}{1+\eta_5}.
\label{defkappa}
\end{equation}
Integrating (\ref{stdismarguqw}), we get, for any $\alpha \ge 0$,
\begin{equation}
\frac{\langle |U|^\alpha \rangle}{\eta_6^\alpha}
=
\frac{F(\kappa_1,\kappa_2,\alpha)}{F(\kappa_1,\kappa_2,0)} \, ,
\label{UmomN1twopar}
\end{equation}
where function $F(\kappa_1,\kappa_2,\alpha)$ is defined by
\begin{eqnarray}
F(\kappa_1,\kappa_2,\alpha)
&=&
\left( 
2 \kappa_1 \kappa_2
\right)^{(1-\alpha) \kappa_2}
\exp
\left(
2 \kappa_1 \kappa_2
\right)
\,
\mathrm{\Gamma} 
\big(
1+(\alpha-1) \kappa_2, 2 \kappa_1 \kappa_2
\big)
\nonumber
\\
&+&
\kappa_1^{(1-\alpha)/2}
\exp(\kappa_1)
{\hskip 0.8mm}
\gamma
\bigg( 
\frac{\alpha+1}{2}, 
\kappa_1
\!\bigg)
\label{deffunfk1k2al}
\end{eqnarray}
and $\mathrm{\Gamma}$ (resp. $\gamma$) is the upper (resp. lower) incomplete gamma
function defined by
$$
\mathrm{\Gamma}(s,y) = \int_y^\infty \xi^{s-1} \exp(-\xi) \, \mbox{d} \xi,
\qquad
\gamma(s,y) = \int_0^y \xi^{s-1} \exp(-\xi) \, \mbox{d} \xi.
$$
Substituting $\alpha=2$ and $\alpha=4$ in equation~(\ref{UmomN1twopar}), we get
\begin{equation}
\mathrm{Kurt}[U]
=
\frac{\langle U^4 \rangle}{\langle U^2 \rangle^2}
=
\frac{F(\kappa_1,\kappa_2,4) \, F(\kappa_1,\kappa_2,0)}{
(F(\kappa_1,\kappa_2,2))^2} \, .
\label{kurtosistwopar}
\end{equation}
This formula for the kurtosis is visualized in Figure~\ref{figure2}(a) as a function
of parameter $\eta_5$ for three different values of parameter $\eta_6$. We note that 
the case $\eta_6=0$ corresponds to the case studied in Section~\ref{oneparam}. 
If $\eta_6=0$, then equation~(\ref{defkappa}) implies $\kappa_1=0$. Since 
$\gamma(s,0)=0$ and
$\mathrm{\Gamma}(s,0)=\mathrm{\Gamma}(s)$, where $\mathrm{\Gamma}(s)$ is the standard gamma function
given by (\ref{gammadef}), we can confirm that equation~(\ref{kurtosistwopar})
converges to our previous result~(\ref{kurtosis}) as $\eta_6 \to 0.$

Substituting $\alpha=1$ into (\ref{deffunfk1k2al}), we obtain
$F(\kappa_1,\kappa_2,1)
=
\exp
\left(
\kappa_1
\right).
$
Consequently, using $\alpha=1$ in equation~(\ref{UmomN1twopar}), we obtain
\begin{equation}
\frac{\langle |U| \rangle}{\eta_6}
=
\frac{\exp \left(\kappa_1\right)}{F(\kappa_1,\kappa_2,0)} \, .
\label{Ualph1twopar2}
\end{equation}
Using $\alpha=2$ in equation~(\ref{UmomN1twopar}), we get
\begin{equation}
\frac{\langle U^2 \rangle}{\langle |U| \rangle^2}
=
\frac{F(\kappa_1,\kappa_2,2) \, F(\kappa_1,\kappa_2,0)}{\exp(2\kappa_1)} \, .
\label{secondtwoparequation}
\end{equation}
Consequently, if we use MD simulations to estimate not only the second and fourth moments,
$\langle U^2 \rangle$ and $\langle U^4 \rangle$, but also the first absolute moment
$\langle |U| \rangle$, we can substitute the estimated MD values into equations
(\ref{kurtosistwopar}) and (\ref{secondtwoparequation}) to obtain two equations for
two unknowns $\kappa_1$ and $\kappa_2$. Solving these two equations numerically,
we can get $\kappa_1$ and $\kappa_2$. Then we can use (\ref{defkappa}) and
(\ref{Ualph1twopar2}) to get the original parameters $\eta_5$ and $\eta_6$ by
\begin{equation}
\eta_5
=
\frac{1-\kappa_2}{\kappa_2}
\qquad \mbox{and}
\qquad
\eta_6
=
\frac{\langle |U| \rangle \, F(\kappa_1,\kappa_2,0)}{\exp \left(\kappa_1\right)} \, .
\label{transetakappa}
\end{equation}
Moreover, equation (\ref{defkappa}) also implies the following restriction on
other parameters $\eta_2$, $\eta_3$ and $\eta_4$
\begin{equation}
\frac{\eta_4^2}{\eta_2 \eta_3}
=
\frac{1-\kappa_2}{\kappa_1 \, \kappa_2 \, \exp \left( \kappa_1/\kappa_2 \right)}
\Big(
\langle |U| \rangle \, F(\kappa_1,\kappa_2,0)
\Big)^{1/\kappa_2}.
\label{kapparestriction}
\end{equation}
This restriction is equivalent to restriction (\ref{eta5restriction}).
Therefore, assuming again that $D$, $\langle V^2 \rangle$, $\langle Z^2 \rangle$
are obtained from MD simulations and $\eta_4^2/(2 \eta_2 \eta_3)$ is given
by (\ref{kapparestriction}), we can calculate parameters $\eta_1$,
$\eta_2$, $\eta_3$ and $\eta_4$ by equations (\ref{eta12formnl})--(\ref{eta34formnl}).

We note that the two additional parameters $\eta_5$ and $\eta_6$ can be used to
satisfy both equations (\ref{kurtosistwopar}) and (\ref{secondtwoparequation}),
while in Section~\ref{oneparam} we could only use one equation 
(equation (\ref{kurtosis}) for kurtosis) to fit one parameter $\eta_5$. 
However, in the case of one-parameter function (\ref{gdefsim}), we could (instead 
of fitting the kurtosis) match the quantity $\langle U^2 \rangle/\langle |U| \rangle^2$
with MD simulations, i.e. we could replace equation~(\ref{kurtosis}) by 
equation~(\ref{secondtwoparequation}) simplified to the one-parameter
case corresponding to function (\ref{gdefsim}).
Passing to the limit $\eta_6 \to 0$ in equation (\ref{secondtwoparequation}) 
and using Euler's reflection formula, $\mathrm{\Gamma}(1-y) \mathrm{\Gamma}(y) \sin(\pi y) = \pi$, 
we obtain that the one-parameter nonlinearity (\ref{gdefsim}) implies the following
formula
\begin{equation}
\frac{\langle U^2 \rangle}{\langle |U| \rangle^2}
=
\frac{\pi}{1+\eta_5}
\left(
\sin
\left(
\frac{\pi}{1+\eta_5}
\right)
\right)^{\!\! -1}
\, .
\label{secondtwoparequatione5}
\end{equation}
Thus, in Section~\ref{oneparam}, we could use $\langle |U| \rangle$ and $\langle U^2 \rangle$
estimated from long-time MD simulations to calculate the left hand side 
of equation (\ref{secondtwoparequatione5}), which could then be used to select parameter
$\eta_5$. Other parameters could again be chosen by equations (\ref{eta12formnl})--(\ref{eta34formnl}).

\subsection{Application to MD simulations}

In Sections~\ref{oneparam} and~\ref{twoparam}, we have presented three approaches to fit 
nonlinear SCG models which have non-Gaussian force distributions to data obtained from MD 
simulations. In this section, we apply them to the results obtained by an illustrative 
MD simulation of a Lennard-Jones fluid, where we consider a box of 512 atoms which
interact with each other through the Lennard-Jones force terms for parameters given
for liquid argon~\citep{Rahman:1964:CMA}, i.e. particles interact in pairs according to 
the Lennard-Jones potential $4 {\hskip 0.2mm} \varepsilon \, ( (\sigma/r)^{12} - (\sigma/r)^6 )$, 
where $\varepsilon/k_B = 120\,$K, $\sigma = 0.34\,$nm and $r$ being the distance 
between particles. We use standard NVT simulations where the temperature ($T=94.4\,$K) 
is controlled using the thermostat of~\cite{Nose:1984:UFC} and \cite{Hoover:1985:CDE} 
and the number of particles ($N=512$ in a cubic box of side $2.91\,$nm) is kept 
constant by implementing periodic boundary conditions. 

Using a long time MD simulation (time series of lentgh 10$\,$ns), 
we estimate three moments $\langle |U| \rangle$, $\langle U^2 \rangle$ and $\langle U^4 \rangle$ 
as averages over all three coordinates, i.e.
$$
\langle |U|^\alpha \rangle
=
\frac{\langle |U_1|^\alpha \rangle + \langle |U_2|^\alpha \rangle + \langle |U_3|^\alpha \rangle}{3},
\qquad
\quad
\mbox{for}
\quad
\alpha=1, 2
\;
\mbox{and}
\;
4,
$$
where $(U_{1},U_{2},U_{3})$ is the acceleration of one specific 
atom (tagged particle) to which our SCG model is applied. Rounding
all computational results to three significant figures, we obtain
$\langle |U| \rangle = 0.753\,$nm$\,$ps$^{-2}$,
$\langle U^2 \rangle = 1.10\,$nm$^2\,$ps$^{-4}$
and
$\langle U^4 \rangle = 7.03\,$nm$^4\,$ps$^{-8}$.

In Figure~\ref{figure2}(b), we plot the equilibrium MD distribution 
of the acceleration (average over all three coordinates) using 
blue circles. The resulting distribution is leptokurtic (with positive excess 
kurtosis). Its kurtosis has been estimated as $\mathrm{Kurt}[U]=5.85$.
The numerical values on the $u$-axis in Figure~\ref{figure2}(b)
are expressed in [nm ps$^{-2}$]. Since the acceleration, $U$, is proportional 
to the force exerted on the tagged particle (with the scaling factor equal 
to the atomic mass of argon), the plot of the acceleration distribution 
in Figure~\ref{figure2}(b) can also be interpreted as the plot of the 
force distribution, which has the same kurtosis, provided that we 
suitably rescale the units on the $u$-axis. 

If we only attempt to fit the value of $\langle U^2 \rangle$, we could 
parametrize the linear SCG model~(\ref{scm1simsim})--(\ref{scm4simsim}), 
which leads to the Gaussian acceleration distribution
(plotted as the black dotted line in Figure~\ref{figure2}(b)). Using the one-parameter nonlinear
function (\ref{gdefsim}) from Section~\ref{oneparam}, we can use equation~(\ref{kurtosis})
to find parameter $\eta_5=0.550$ so that the nonlinear SCG model gives the same kurtosis
as observed in MD simulations ($\mathrm{Kurt}[U]=5.85$).
The resulting distribution is given as the red dot-dashed line in Figure~\ref{figure2}(b).
It matches both second and fourth moments, $\langle U^2 \rangle$ and $\langle U^4 \rangle$.

Using all-atom MD simulations, we can not only estimate the kurtosis, but other dimensionless
ratios of moments of $U$. For example, we obtain $\langle U^2 \rangle/\langle |U| \rangle^2=1.93$. 
This estimate can be substituted in equation~(\ref{secondtwoparequatione5}), which provides
an alternative approach to obtain the value of parameter $\eta_5$ of the one-parameter nonlinear
function (\ref{gdefsim}). Using $\langle U^2 \rangle/\langle |U| \rangle^2=1.93$
and solving equation~(\ref{secondtwoparequatione5}) numerically, we obtain $\eta_5=0.692.$
The resulting distribution, which matches $\langle |U| \rangle$ and $\langle U^2 \rangle$,
is plotted as the green dashed line in Figure~\ref{figure2}(b). 
We note that the parameter $\eta_5$ is dimensionless, 
because both equations (\ref{kurtosis}) and (\ref{secondtwoparequatione5}) only depend on
dimensionless quantities estimated from MD simulations. Since both distributions (for
$\eta_5=0.550$ and $\eta_5=0.692$) are given by~(\ref{stdisN1nonlin2}), they are
unbounded for $u$ close to zero. This motivates the choice of our two-parameter 
nonlinear function $g$ used in Section~\ref{twoparam}. 

Substituting $\mathrm{Kurt}[U]=5.85$ and $\langle U^2 \rangle/\langle |U| \rangle^2=1.93$
in equations~(\ref{kurtosistwopar}) and~(\ref{secondtwoparequation}) and solving them
numerically, we obtain $\kappa_1=0.149$ and $\kappa_2=0.771$. 
Substituting into~(\ref{transetakappa}), we get the two parameters of
model from Section~\ref{twoparam} as $\eta_5=0.297$ and $\eta_6=0.472$ nm$\,$ps$^{-2}$. 
The resulting distribution, given by equation~(\ref{stdismarguqw}), 
is plotted in Figure~\ref{figure2}(b) as the cyan
solid line. We observe that the distribution is now bounded. It is 
a piecewise defined function which is Gaussian for the values of $u$ 
satisfying $|u| \le \eta_6$, which removes the singularity at $u=0$.
At the same time, the distribution given by equation~(\ref{stdismarguqw})
matches all three moments estimated from MD simulations,
$\langle |U| \rangle$, $\langle U^2 \rangle$ and $\langle U^4 \rangle$.
As we can see in Figure~\ref{figure2}(b), this distribution
does not perfectly fit the acceleration distribution estimated from MD
simulations. If our aim is to obtain a SCG model which better fits the
whole distribution, we can use SCG models for larger values of $N$ as 
we will discuss in the next section.

\section{Nolinear SCG model for general values of $N$}

\label{secgenNnonlin}

We have already observed in Sections~\ref{secN1lin} and~\ref{secgenNlin}
that the linear SCG model (\ref{scm1sim})--(\ref{scm4sim}) can match the MD
values of a few moments for $N=1$, while we need to consider larger values 
of $N$ to match the entire velocity autocorrelation function. Considering
the nonlinear SCG model~(\ref{scm1})--(\ref{scm4}), we have two options
to capture more details of the non-Gaussian force distribution observed 
in MD simulations. We could either keep $N=1$, as in Section~\ref{secN1nonlin},
and introduce additional 
parameters into nonlinearity $g=g_1$, or we could consider larger values of $N$.  
In Section~\ref{secN1nonlin}, we have shown that by going from one-parameter to
two-parameter function $g$, we improve the match with MD results. In this
section, we will discuss the second option: we will use larger values of $N$.
 
Consider equations corresponding to the $i$-coordinate, $i=1$, $2$, $3$, of the nonlinear 
SCG model~(\ref{scm1})--(\ref{scm4}). Let us denote the stationary distribution
of equations~(\ref{scm2})--(\ref{scm4}) by 
$$
p(v,{\mathbf u},{\mathbf z}) 
\equiv
p(v,u_1,u_2,\dots,u_N,z_1,z_2,\dots,z_N).
$$
Then 
$p(v,{\mathbf u},{\mathbf z}) \, \mbox{d}v \, \mbox{d}u_1 \, \mbox{d}u_2 \, \dots \, \mbox{d}u_N 
\, \mbox{d}z_1 \, \mbox{d}z_2 \, \dots \, \mbox{d}z_N$ 
gives the probability that $V_i(t) \in [v,v+\mbox{d}v)$,
$U_{j,i}(t) \in [u_j,u_j+\mbox{d}u_j)$ and
$Z_{j,i}(t) \in [z_j,z_j+\mbox{d}z_j)$, for $j=1$, $2$, $\dots$, $N$, at equilibrium.
The stationary distribution can be obtained by solving the 
corresponding stationary Fokker-Planck equation
\begin{eqnarray}
\frac{\eta_{j,4}^2}{2}
\frac{\partial^2 p}{\partial^2 z_j}
(v,{\mathbf u},{\mathbf z})
&=&
\frac{\partial}{\partial v}
\left(
p(v,{\mathbf u},{\mathbf z}) \, \sum_{j=1}^{N} u_j
\right)
\nonumber
\\
&+&
\sum_{j=1}^{N}
\frac{\partial}{\partial u_j}
\Big(
\big(
-
\eta_{j,1} v
+
h_j(z_j)
\big)
g_j^\prime(g_j^{-1}(u_j)) \,
p(v,{\mathbf u},{\mathbf z})
\Big)
\nonumber
\\
&+&
\sum_{j=1}^{N}
\frac{\partial}{\partial z_j}
\Big(
\big(
- \eta_{j,2} h_j(z_j) 
- \eta_{j,3} u_j
\big)
p(v,{\mathbf u},{\mathbf z})
\Big) \, .
\label{statfp}
\end{eqnarray}
Our analysis in Section~\ref{oneparam} shows that
parameters $\eta_{j,2}$, $\eta_{j,3}$ and $\eta_{j,4}$
appear on the left hand side of equation~(\ref{eta5restriction})
as a suitable fraction,
which in the Gaussian case corresponds to the second moment of the
acceleration (see equation~(\ref{gaussianrestriction})). Considering
general $N$, we define this fraction as new parameters 
$$
\sigma_j 
=
\frac{\eta_{j,4}^2}{2 \, \eta_{j,2} \, \eta_{j,3}} \, ,
\qquad\quad \mbox{for} \qquad j=1,2,\dots,N,
$$
and we again assume that the second moment of the velocity distribution,
$\langle V^2 \rangle = \langle V_i^2 \rangle$, can be estimated from 
long-time MD simulations. In order to find the stationary distribution, 
we will require that parameters 
$\eta_{j,1}$, $\eta_{j,2}$, $\eta_{j,3}$ and $\eta_{j,4}$
satisfy (compare with equation (\ref{VZsecmom}) for $N=1$)
$$
\langle V^2 \rangle
=
\frac{\eta_{j,4}^2}{2 \, \eta_{j,1} \, \eta_{j,2} \, \eta_{j,3}} 
=
\frac{\sigma_j}{\eta_{j,1}} \,,
\qquad\quad  \mbox{for all} \qquad j=1,2,\dots,N.
$$
Then the stationary distribution, obtained by solving (\ref{statfp}), 
is given by
\begin{eqnarray}
p(v,{\mathbf u},{\mathbf z})
=
C
\,
\left(
\prod_{j=1}^N
\frac{1}{g_j^\prime(g_j^{-1}(u_j))}
\right)
\,
\exp \!
\Bigg[
&&
\!\!\!
- \, \frac{v^2}{2 \, \langle V^2 \rangle}
-
\sum_{j=1}^N
\frac{1}{\sigma_j} \, G_j \big( g_j^{-1}(u_j) \big)
\nonumber
\\
&&-
\sum_{j=1}^N
\frac{2 \eta_{j,2}}{\eta_{j,4}^2} 
H_j(z_j)
\Bigg],
\label{stdistN}
\end{eqnarray}
where $C$ is the normalization constant and functions $G_j$ and $H_j$ are
integrals of functions $g_j$ and $h_j$, respectively, which are given by
\begin{equation*}
G_j(y) = \int_0^y g_j(\xi) \, \dxi \, ,
\qquad
H_j(y) = \int_0^y h_j(\xi) \, \dxi \,,
\qquad  \mbox{for} \quad j=1,2,\dots,N.
\end{equation*}
Following (\ref{gdefsim}), we assume that $h_j(z_j)=z_j$
and each $g_j$ is a function of one additional positive
parameter $\eta_{j,5},$ $j=1$, $2$, $\dots$, $N$,
given as
\begin{equation}
g_j(y) 
= 
\left|
y
\right|^{1/\eta_{j,5}}
\sign y \, .
\label{gjdefoneparam}
\end{equation}
Then we have
\begin{equation*}
g_j^\prime(g_j^{-1}(u_j)) 
= 
\frac{
\left|
u_j
\right|^{1-\eta_{j,5}}}{\eta_{j,5}}
\qquad
\mbox{and}
\qquad
G_j \big( g_j^{-1}(u_j) \big)
= 
\frac{\eta_{j,5}}{1+\eta_{j,5}} 
|u_j|^{1+\eta_{j,5}}.
\end{equation*}
Then the stationary distribution (\ref{stdistN}) is Gaussian in $V_i$ and
$Z_{j,i}$ variables and we can integrate (\ref{stdistN}) to calculate the 
marginal distribution of $U_{j,i}$ by
$$
p_j(u_j)
=
\int_{-\infty}^\infty
\dots
\int_{-\infty}^\infty
p(v,{\mathbf u},{\mathbf z}) \, \dv \,
\du_1 \, \du_2 \, \dots \, \du_{j-1} \, \du_{j+1} \, \dots \, \du_N 
\, \mbox{d} {\mathbf z} \,.
$$
Consequently,
\begin{equation}
p_j(u_j)
=
C_j
|u_j|^{\eta_{j,5}-1}
\,
\exp
\left[
-
\frac{\eta_{j,5}}{\sigma_j (1+\eta_{j,5})} 
|u_j|^{1+\eta_{j,5}}
\right],
\label{stdismargju}
\end{equation}
where the normalization constant $C_j$ is given by
$$
\int_{-\infty}^\infty
p_j(u_j) \, \du_j
=
1.
$$
Integrating (\ref{stdismargju}), we can calculate
$$
\langle |U_{j,i}|^\alpha \rangle
=
\int_{-\infty}^\infty
|u_j|^\alpha
p_j(u_j) \, \du_j, \qquad \mbox{for any} \; \alpha \ge 0,
$$
as
\begin{equation}
\langle |U_{j,i}|^\alpha \rangle
=
\left(
\frac{\sigma_j(1+\eta_{j,5})}{\eta_{j,5}}
\right)^{\alpha/(1+\eta_{j,5})}
\frac{
\mathrm{\Gamma} \!
\left(
\frac{\alpha+\eta_{j,5}}{1+\eta_{j,5}}
\right)
}{
\mathrm{\Gamma} \!
\left(
\frac{\eta_{j,5}}{1+\eta_{j,5}}
\right)
}.
\label{absmomgj2}
\end{equation}
The acceleration of the coarse-grained particle is given by 
\begin{equation*}
U_{i} = \sum_{j=1}^N U_{j,i}.
\end{equation*}
Using the symmetry of (\ref{stdismargju}), odd moments of $U_{j,i}$
are equal to zero. In particular, $\langle U_{j,i} \rangle = 0$
and $\langle U_{j,i}^3 \rangle = 0$ for $j=1$, $2$, $\dots$, $N$.
Consequently,
\begin{eqnarray}
\langle U_{i}^2 \rangle
&=&
\sum_{j=1}^N
\langle U_{j,i}^2 \rangle \, ,
\label{secmomgenN}
\\
\langle U_{i}^4 \rangle
&=&
3
\langle U_{i}^2 \rangle^2
+
\sum_{j=1}^N
\langle U_{j,i}^4 \rangle
-
3
\langle U_{j,i}^2 \rangle^2
\, ,
\label{fourmomgenN}
\end{eqnarray}
which gives
\begin{equation}
\mathrm{Kurt}[U_i]
=
\frac{\langle U_i^4 \rangle}{\langle U_i^2 \rangle^2}
=
3
+
\frac{
\sum_{j=1}^N
\langle U_{j,i}^4 \rangle
-
3
\langle U_{j,i}^2 \rangle^2
}{
\sum_{j=1}^N
\langle U_{j,i}^2 \rangle} \, .
\label{kurtosisgeneral}
\end{equation}
Substituting equation (\ref{absmomgj2}) for moments on the right hand side of 
equation~(\ref{kurtosisgeneral}), we can express the kurtosis of $U_i$ in terms
of $2N$ parameters $\sigma_j$ and $\eta_{j,5}$, where $j=1$, $2$, $\dots$, $N$. 
For example, if we choose the values of dimensionless parameters 
$\eta_{j,5}$ equal to given numbers and define new parameters
$$
\kappa_j = 
\big(
\sigma_j
\big)^{2/(1+\eta_{j,5})},
$$
then equation (\ref{absmomgj2}) implies that $\langle U_{j,i}^2 \rangle$
is a linear function of $\kappa_j$ and $\langle U_{j,i}^4 \rangle$
is a quadratic function of $\kappa_j$. Equations (\ref{secmomgenN})
and (\ref{fourmomgenN}) can then be rewritten as
the following system of two equations for
$\kappa_1,$ $\kappa_2,$ $\dots,$ $\kappa_N$
$$
\sum_{i=1}^N c_{1,j} \kappa_j
=
\langle U_{i}^2 \rangle,
\qquad
\sum_{i=1}^N c_{2,j} \kappa_j^2
= 
\langle U_{i}^4 \rangle - 3 \langle U_{i}^2 \rangle^2,
$$
where $c_{1,j}$ and $c_{2,j}$ are known constants, which will depend on our initial
choice of values of $\eta_{j,5}$. Thus, using $N>2$, we still have an opportunity 
to not only fit the second and fourth moments of the force distribution, but other 
moments as well. For example, the $6$-th moment, $\langle U_{i}^6 \rangle$, would 
include the linear combination of the third powers of $\kappa_j$. 
We could also fit other properties of the force distribution estimated from 
MD simulations. For example, we could generalize one-parameter
nonlinearities~(\ref{gjdefoneparam}) to two-parameter nonlinear 
functions, as we did in equation~(\ref{gdef2an578qw}). Then we could 
match the value of the distribution at $u=0$, if our aim was to 
get a better fit of the MD acceleration distribution obtained 
in the illustrative example in Figure~\ref{figure2}(b). Another possible
generalization is to consider nonlinear functions $h_j$, provided 
that we estimate more statistics on the auxiliary variable $Z$ from
MD simulations.

\section{Discussion and conclusions}

We have presented and analyzed a family of SCG models given by 
equations~(\ref{scm1})--(\ref{scm4}), which can be parametrized
to fit properties of detailed all-atom MD models. A special choice
of functions $g_j$ and $h_j$ in equations~(\ref{scm1})--(\ref{scm4}) 
leads to the linear SCG model~(\ref{scm1sim})--(\ref{scm4sim}) 
which is used in a multiscale (multi-resolution) method 
developed by~\citet{Erban:2016:CAM} as an intermediate 
description between all-atom MD simulations and BD models. 
The linear SCG model is studied in more detail 
in Sections~\ref{secN1lin} and \ref{secgenNlin}, where we highlight
that $4N$ parameters of this model can match some statistics
estimated from all-atom MD simulations with increased accuracy as
we increase $N$, but there are also statistics which cannot be
matched for any value of $N$. They include non-Gaussian force
distributions.

In Sections~\ref{secN1lin} and \ref{secgenNlin}, we show that the
linear SCG model~(\ref{scm1sim})--(\ref{scm4sim}) corresponds to
the generalized Langevin equation with the stochastic driving
force being Gaussian. Such systems have been analysed since 
the work of~\citet{Kubo:1966:FDT}. One approach to match non-Gaussian
MD force distributions could be to use the non-Gaussian generalized 
Langevin equation which was analyzed by~\citet{Fox:1977:ANG} using 
methods of multiplicative stochastic processes. However, if we 
want to generalize the linear SCG model~(\ref{scm1sim})--(\ref{scm4sim})
while keeping its structure as a relatively low-dimensional system
of SDEs, then it can be done by
introducing nonlinear functions $g_j$ and $h_j$ as shown in 
equations~(\ref{scm1})--(\ref{scm4}). The advantage of the
presented approach
is that we can directly replace the linear model by
equations~(\ref{scm1})--(\ref{scm4}) in multiscale methods which use 
all-atom MD simulations in parts of the computational 
domain and (less detailed) BD simulations in the remainder of the domain.
Coupling MD and BD models is a possible approach to incorporate
atomic-level information into models of intracellular processses 
which include transport of molecules between different parts of
the cell~\citep{Erban:2014:MDB,Erban:2016:CAM,Gunaratne:2019:MDM}.

The nonlinear SCG model~(\ref{scm1})--(\ref{scm4}) is studied in
Section~\ref{secN1nonlin} for $N=1$. Describing the nonlinearity
as the one-parameter function given by~(\ref{gdefsim}), 
we can use its dimensionless parameter $\eta_5$ to match 
the kurtosis of the force distribution estimated from all-atom 
MD simulations. Although the one-parameter case is
easy to analyze in terms of the gamma function, it has some undesirable
properties for small forces. If $\eta_5>1$, we can obtain large terms in
the dynamical equation~(\ref{scm3nonlins}) for small values of $U$;
this corresponds to the zero value of stationary 
probability distribution~(\ref{stdisN1nonlin2}) for $u=0$.
If $\eta_5<1$, we have small terms in the dynamical 
equation~(\ref{scm3nonlins}), but the stationary probability 
distribution~(\ref{stdisN1nonlin2}) is unbounded for 
$u=0$. In Section~\ref{twoparam}, we show that these issues can be
avoided if the two-parameter nonlinear function~(\ref{gdef2an578qw})
is used instead of the one-parameter function~(\ref{gdefsim}). 
The resulting equations are solved in terms of incomplete gamma functions. 
In Section~\ref{secgenNnonlin}, we study the nonlinear 
model for general values of $N$ where each
$g_j$ is a one-parameter nonlinearity given by equation~(\ref{gjdefoneparam}). 
However, we could also consider two-parameter functions $g_j$, like
we did in equation~(\ref{gdef2an578qw}) for $N=1$, to improve the properties of the 
SCG model for general values of $N$.

\newpage

\noindent {\bf Acknowledgements.}

\noindent
I would like to thank the Royal Society for a University Research Fellowship.

\end{document}